\documentclass{ws-p8-50x6-00}

\begin{document}

\title{ Saturation in DIS processes
}

\author{Errol Gotsman}

\address{[in collaboration with E. Levin, U. Maor and E. Naftali]}  

\address{School of Physics and Astronomy, Tel Aviv University,  
 Tel Aviv 69978 Israel\\ 
E-mail: gotsman@post.tau.ac.il }


\maketitle

\abstracts{ We examine HERA data with a view of determining
 whether unique signs of saturation can be identified. Concentrating on
two channels the logarithmic slope of $F_{2}$, and the production of
$J/\Psi$,
 which are sensitive to the behaviour of $xG(x,Q^{2})$ the gluon density
distribution in the proton,
 we show that our model incorporating screening corrections and 
alternative models comprising a sum of a "soft" and "hard"
component provide good fits to the data.}

 \section{Introduction}

\par The problem  we  address, is whether the 
 experimental results eminating from HERA \cite{exp1} contain
 clear  evidence for the presence of saturation
effects, or whether they are consistent with orthodox pQCD evolution.

\par To quantify saturation it is instructive to introduce the concept of
a packing factor (PF) which is related to the density of the partons in a
parton
cascade.
\begin{equation} \label{EQ2}
PF \,\equiv\,\kappa\,\,\,=\,\,\,\frac{3\,\pi^2 \alpha_S}{2
Q^2_s(x)}\,\times\,
\frac{xG(x,Q^2_s(x))}{\pi\, R^2}\,\,
\end{equation}The saturation scale $Q^{2}_{s}$ is defined by $ \kappa\, = \,1$, for
which
 $Q^{2}\, = \,Q^{2}_{s}$.
(see fig.1)

\begin{figure}
\epsfig{file=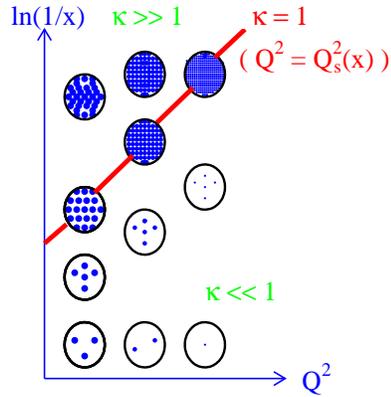,width=6cm,height=6cm}
\caption{ Parton distribution in the transverse plane.
\label{Fig.1}}
\end{figure}

In the dilute region ($\kappa \, < \, 1$) the partons are distant from
one another (and have no interaction in the parton cascade), so pQCD
(i.e.  DGLAP and BFKL)  evolution holds, and the dominant process is the
emission of gluons. In the high density phase ($\kappa \, > \, 1$) the
partons in the parton cascade interact, these interactions
give rise to screening corrections (SC), which slow down the
growth of the parton density distributions. The correct description of
parton evolution in the high density phase is given by a non-linear
evolution equation of the type first suggested by Gribov, Levin and Ryskin
\cite{GRL}, which  incorporates parton recombonation as well as
emission.    

 \par There have been numerous attempts to find both approximate
analytical and numerical solutions to the non-linear equation (for a
recent review see \cite{Lub}).  These solutions   suggest that
the saturation scale $Q_{S}(x)\, \approx \, 1-2 \, GeV^{2}$, in the HERA
kinematic region. The very successful phenomenological model of
Golec-Biernat and Wuestoff
\cite{GB-W}, which  provides an excellent description of HERA
data based on the premise that the saturation region has been reached at
HERA, is additional evidence supporting the saturation hypothesis.

\section{ The GLMN model}
  In a series of papers, the latest of which are
listed in  \cite{PL2001a},
  we have applied screening 
(unitarity) corrections  to a number of inclusive and exclusive
channels.
We follow the Glauber-Mueller (eikonal) 
approach
\cite {ZKL} in calculating screening (unitarity) corrections to  pQCD
evolution.
 The technique used for
evaluating the 
SC in the quark and gluon sectors are  given
 in reference \cite{GLM1}.

 As the SC are much larger for the gluon sector than for the quark sector,
we quote results here for two channels that in LLA of pQCD are directly
proportional to $xG(x,Q^{2})$, the gluon distribution in the proton.

1) The logarithmic slope of $F_{2}$:
 \begin{equation}  
\frac{\partial F_2(x,Q^2)}{\partial \ln Q^2}\,=
\,\frac{2\alpha_S}{9\pi}xG^{DGLAP}(x,Q^2),
\end{equation}
2) The cross section for the exclusive production of the vector meson
$J/\Psi$. For which
the contribution of
pQCD to the imaginary part of the $t=0$ differential cross
section
is given by
\begin{center}
$$
(\frac{d\sigma(\gamma^{*} p \rightarrow V p)}{dt}\,)_{t=0}^{pQCD}\,=\,
\frac{\pi^{3}\Gamma_{ee}M_V^{3}}{48\alpha}\,
\frac{\alpha_{S}^{2}({\bar Q}^{2})}{{\bar Q}^{8}}\,
\,(xG^{DGLAP}(x,{\bar Q}^{2})\,)^{2} \,\,(1\,+\,\frac{Q^{2}}{M_V^{2}}\,),
$$
\end{center}

here  $xG^{DGLAP}$ denotes the gluon distribution function as obtained
from
the DGLAP analysis. 

Our results. see Fig.2 and Fig.3, suggest that
already at HERA energies the SC are considerable
for tse two channels.
 
%
\begin{figure}[hptb]
\begin{tabular}{c c l }
\epsfig{file=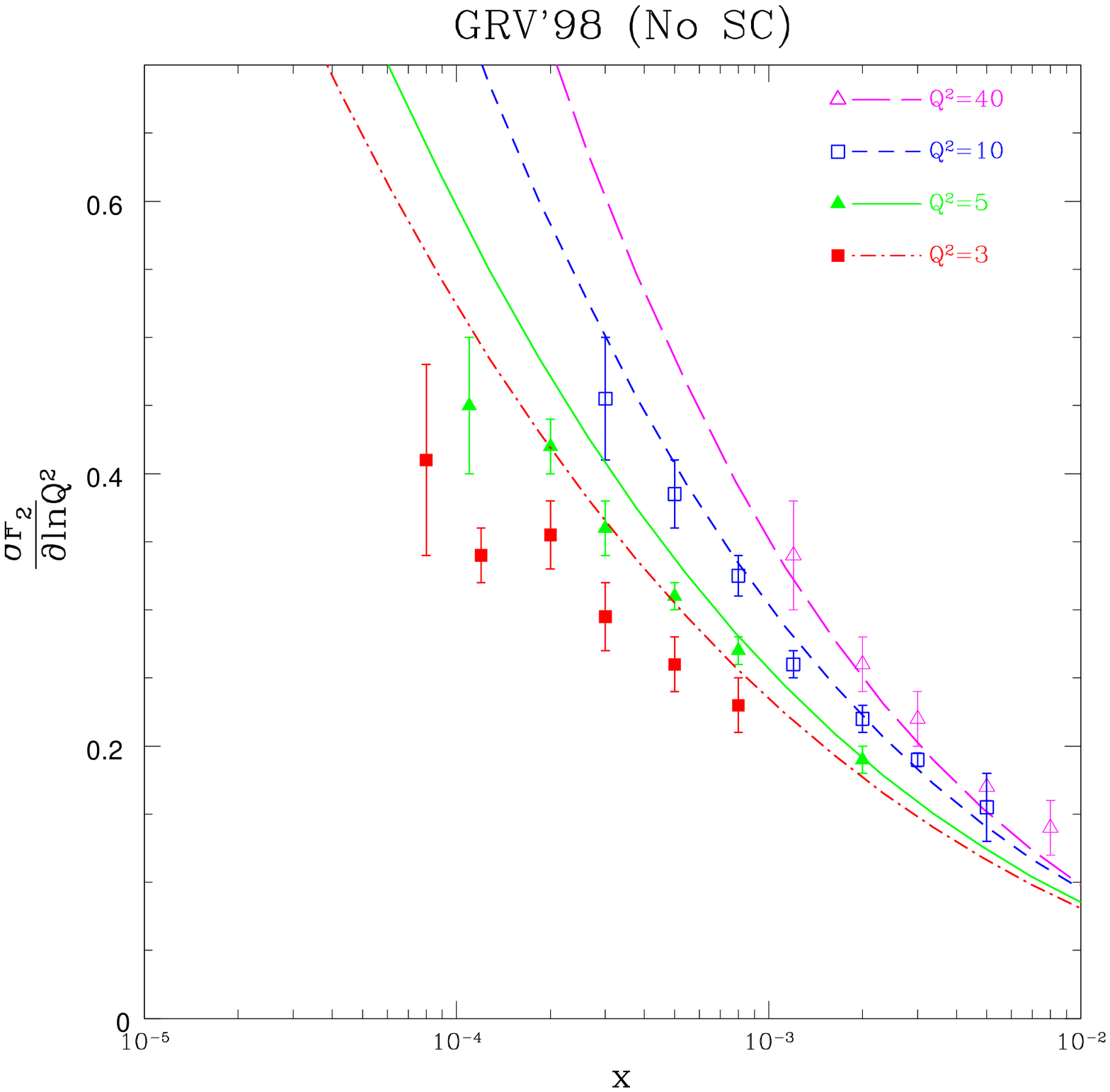,width=56mm,height=6cm} &
 \epsfig{file=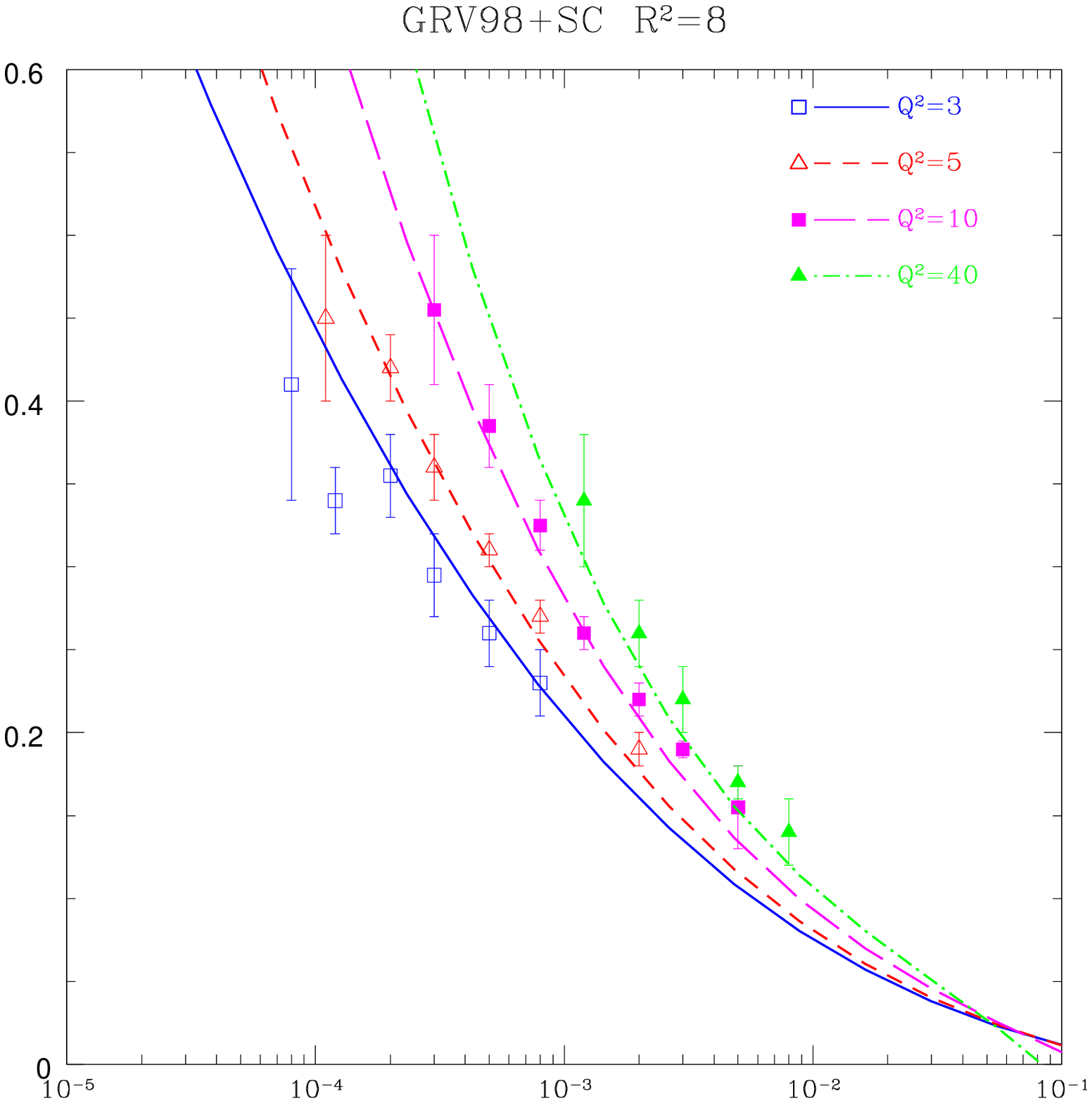,width=56mm,height=6cm}\\
\end{tabular}
\caption{Logarithmic derivative of $F_{2}$ without and with SC.
\label{Fig.2}}
\end{figure}

\begin{figure}[h]
\epsfig{file=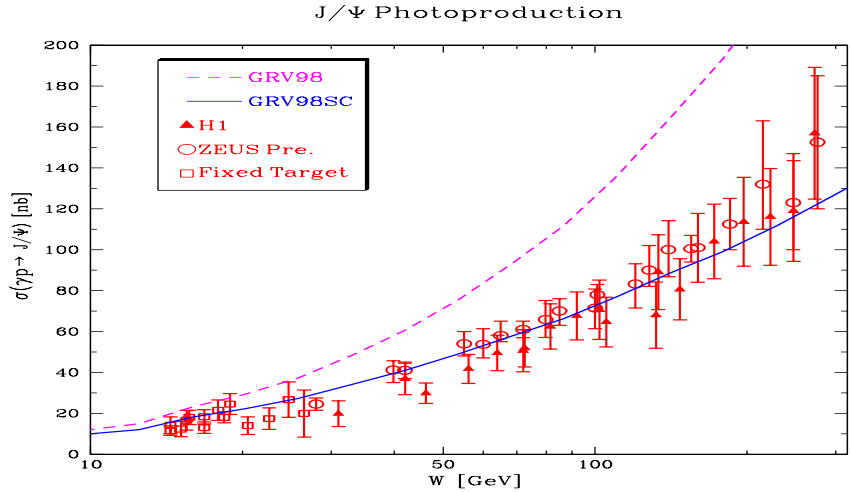,width=12cm,height=7cm}
\caption{$J/\Psi$ photo production with and without SC.
\label{Fig.3}}
\end{figure}

\section{Alternate Models}
   Although, the results presented above are consistent with the
hypothesis
that signs of saturation have been seen at HERA, they are not
conclusive. Since models based on the sum of the contributions  of a
"hard" and  a soft "Pomeron" e.g.
\cite {DL} and \cite {FKS}, provide a fair description of the HERA data.
These models to not incorporate pQCD evolution, but have a common feature
in that they both require the "soft" Pomeron component  to be
appreciable at fairly small scales $\approx$ 0.3 - 0.5 fm.
\begin{figure}[h]
\epsfig{file=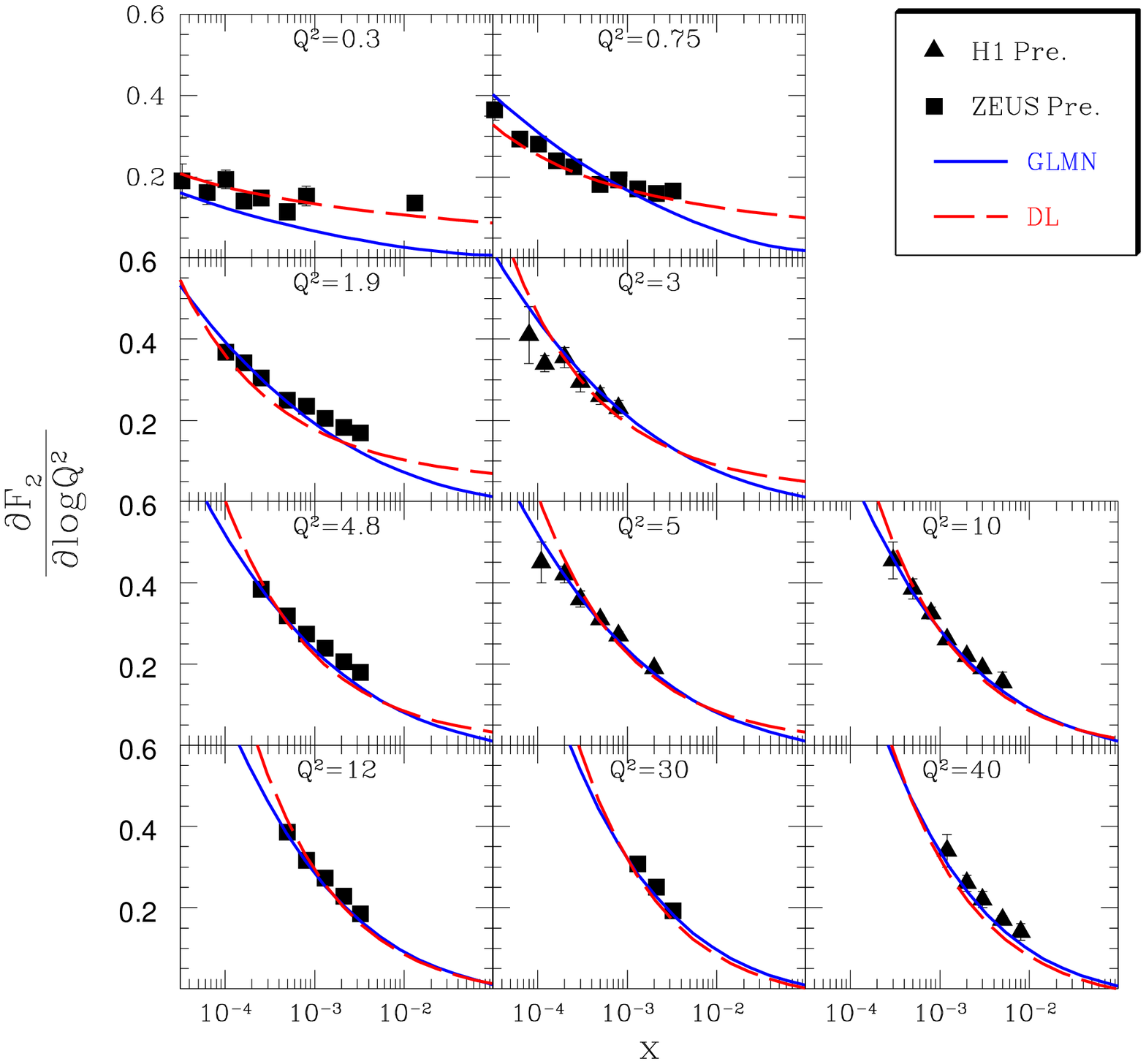,height=8cm,width=12cm}
\caption{ W dependence of HERA data for logarithmic slope at fixed
  $Q^{2}$ (in $GeV^{2}$) compared with our calculations for screened
  GRV98 and the DL model.
\label{Fig.4}}
\end{figure}

 In Fig.4 we compare the DL predictions for 
$\frac{\partial F_2(x,Q^2)}{\partial \ln Q^2}$ with those of our model
(GLMN) i.e. screened GRV'98, and show that there is little to choose
between them for $Q^{2} \;  \geq \; 1.9 \; GeV^{2}$. For  values of
 $Q^{2}\; \leq \;  1 \; GeV^{2}$  there is
no justification for using pQCD (our model).

\section{Conclusions}
  Although, HERA data is consistent with the hypothesis
 that we are dealing with parton densities
that are sufficiently dense ($\kappa \, \approx \, 1$), that SC are
necessary. The findings are not conclusive as an alternative explanation
is also valid i.e.
that of a matching between a "soft" and a "hard" process (e.g. the DL
model) where the "soft" component dominates at
relatively short distances of 0.3 - 0.5 fm. 


\end{document}